\documentclass[11pt]{article}
\usepackage{amsmath,amssymb,fullpage,graphicx}
\usepackage{algorithm,algpseudocode}
\usepackage{caption,subcaption}
\usepackage{hyperref}
\usepackage{natbib}
\usepackage{stackengine}
\usepackage{soul}
\usepackage{xcolor}
\usepackage{halloweenmath}

\usepackage{mathtools}
\mathtoolsset{showonlyrefs}

\let\hat\widehat
\let\bar\overline

\newtheorem{theorem}{Theorem}

\newtheorem{proposition}[theorem]{Proposition}
\newtheorem{remark}[theorem]{Remark}

\newenvironment{proof}{{\bf Proof.}}{$\Box$}

\usepackage{sectsty}
\allsectionsfont{\bfseries\sffamily}

\usepackage{tikz}

\usetikzlibrary{positioning,shapes.geometric,calc,
shapes,arrows.meta,arrows,decorations.markings, external, trees}
\tikzstyle{Arrow} = [
         thick,
         decoration={
                 markings,
                 mark=at position 1 with {
                         \arrow[thick]{latex}
} },
         shorten >= 3pt, preaction = {decorate}
         ]

\def\var{\textnormal{Var}}
\def\cov{\textnormal{Cov}}

\DeclareMathOperator*{\argmin}{argmin}

\newcommand{\E}{\mbox{$\mathbb{E}$}}

\def\adj{\mathrm{(RWR)}}
\def\naive{\mathrm{(naive)}}

\def\Xt{\widetilde{X}}

\def\regression{\textsc{regression}}
\def\i{{(i)}}

\begin{document}
\begin{center}
\textsf{\textbf{\Large 
Nonlinear Regression with Residuals: Causal Estimation with Time-varying
Treatments and Covariates
}}\\
\vspace{.35cm}
Stephen Bates$^1$, Edward Kennedy$^2$, Robert Tibshirani$^3$, 
Val\'erie Ventura$^4$ and Larry Wasserman$^5$ \\

\vspace{.25cm}
{\small
\begin{itemize}
    \itemsep-.3em 
 \item[$^1$] Departments of EECS and Statistics, University of California Berkeley; stephenbates@berkeley.edu
 \item[$^2$] Department of Statistics \& Data Science, Carnegie Mellon University; edward@stat.cmu.edu
  \item[$^3$] Departments of Biomedical Data Science and Statistics, Stanford University; tibs@stanford.edu
  \item[$^4$] Department of Statistics \& Data Science and Neuroscience Institute, Carnegie Mellon University; vventura@stat.cmu.edu
   \item[$^5$] Departments of Statistics \& Data Science and of Machine Learning, Carnegie Mellon University; larry@stat.cmu.edu
\end{itemize}
}

\end{center}
\begin{abstract}
Standard regression adjustment gives inconsistent estimates of causal effects when there are time-varying treatment effects and time-varying covariates. Loosely speaking, the issue is that some
covariates are post-treatment variables because they may be affected by prior treatment status, and regressing out post-treatment variables causes bias.
More precisely, the bias is
due to certain non-confounding latent variables that create colliders in the causal graph.
These latent variables, which we call {\em phantoms}, do not harm the identifiability
of the causal effect, but they render naive regression estimates inconsistent.
Motivated by this, we ask: how can we 
modify regression methods so that they hold up even in the presence of phantoms?
We develop an estimator for this setting based on regression modeling (linear, log-linear, probit and Cox regression), proving that it is consistent for a reasonable causal estimand.
In particular, the estimator is a regression model fit with a simple adjustment for collinearity, making it easy to understand and implement with standard regression software.
The proposed estimators are instances of the parametric g-formula, extending the regression-with-residuals approach to several canonical nonlinear models.
\end{abstract}

\section{Introduction}

Regression adjustment is likely the most commonly-encountered causal inference method: the analyst regresses an outcome variable against treatment variables and observed covariates and then interprets the coefficient of the treatment variables as causal estimates.
But when analyzing longitudinal data with time-varying treatments and covariates, regression methods can lead
to inconsistent estimates of causal effects~\citep{robins1986new}. This is due to latent variables
that affect both the outcome and covariates but are \emph{not} unmeasured confounders. For example, suppose we wish to understand the impact of a medication on survival, and also have
weight and blood pressure as covariates. 
Then, the (latent) overall health of a subject affects both their outcome and covariates, 
but not necessarily their choice to take medication.
We call these unmeasured variables {\it phantoms}, distinguishing them from unmeasured confounders.
Phantoms do not disrupt the identifiability of causal effects, but they cause problems 
for standard regression estimates, biasing them
even in the large-sample limit.
From the point of view of causal graphs, this is because phantoms create colliders---variables with two arrows pointing into them as in Figure~\ref{fig::dag}.
Methods have been developed to avoid this problem, such as
marginal structural models~\citep{robins2000marginal}, structural nested models
\citep{robins2000}, and g-formula methods~\citep{naimi2016}. In particular, regression with residuals is a specific estimator for linear structural nested models that requires only a small number of regression fitting steps~\citep{almirall2010structural}.
Adopting this viewpoint, we study the time-varying treatment problem through the lens of collinearity in regression models. We explain how phantoms invalidate naive estimates and derive a collinearity-adjusted regression estimate that is valid even in the presence of phantoms.


\subsection{Motivating examples}
\label{subsec:teaser}

\begin{figure}
\centering
\includegraphics[width = 6in]{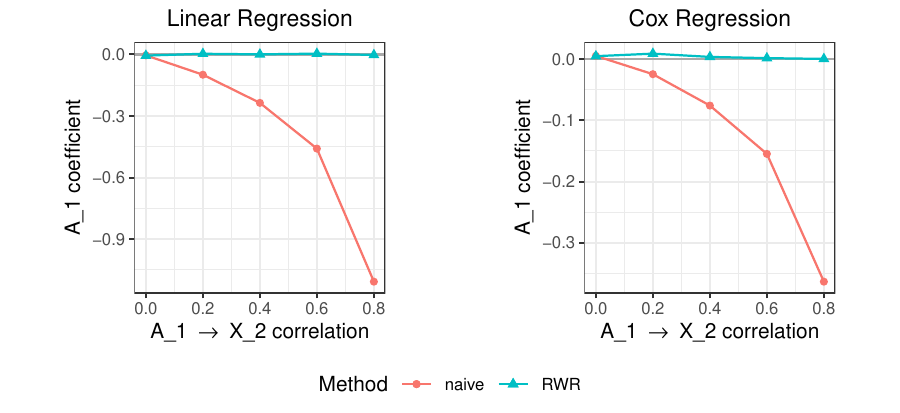}
\caption{\em (Left) Linear regression in a Gaussian model. We plot the regression
coefficient of $A_1$ from naive linear regression and the
proposed method. 
(Right) Cox regression with surival outcomes. We report on the regression coefficient of $A_1$ for naive Cox regression and the proposed method. In both panels, there is no true causal effect (solid horizontal line).
}
\label{fig:teaser}
\end{figure}

To see the shortcomings of standard regression modeling, 
we first consider a simple linear model with Gaussian errors. The outcome is a Gaussian random variable $Y$. We have three observed
explanatory variables: the treatment level at time $1$, $A_1$, 
an observed covariate $X_2$ that is affected by $A_1$, 
and the treatment level at time $2$, 
$A_2$, which is affected by $X_2$. 
There is no unmeasured confounding, but $X_2$ is correlated with $Y$. In the language of this paper, there is a phantom variable---see Section~\ref{sec:setting}. Lastly, there is no treatment effect from $A_1$ or $A_2$ to $Y$. See the right side of Figure~\ref{fig::dag} for a visualization of these dependencies.

With $Y$ Gaussian and mean a linear function of $(A_1, X_2, A_2)$, we generate $n = 100,000$ data points  from the model and then fit a standard linear regression of $Y$ onto $(A_1, X_2, A_2)$. 
The left side of Figure~\ref{fig:teaser} shows the estimated regression coefficient of $A_1$ versus the correlation $\rho$ between $A_1$ and $X_2$.
The estimate deviates farther from zero as $\rho$ increases, which is misleading because there is no causal effect of $A_1$ on $Y$.
By contrast, regression with residuals, a method that will be described in this work,
yields the correct estimate up to simulation error. We will revisit this simple setting again soon in Section~\ref{subsec:phantoms_linear_regression_teaser}, where we derive formulae underlying this simulation result.

The behavior we see in the above is widespread. To demonstrate this,
we carry out a second experiment in a parallel setup, but with survival outcomes.
In this, $Y$ is now a survival time,
but the setting is otherwise unchanged. In particular, $Y$ is correlated with $X_2$, but 
there is again no true causal effect.
We run a Cox regression and 
report the coefficient for $A_1$ in the
right side of Figure~\ref{fig:teaser}, which reveals the and same trend; as the effect from
$A_1$ to $X_2$ increases, the regression coefficient for $A_1$ becomes more
misleading.

\subsection{Related work}
\label{sec:relatedwork}

Causal inference with longitudinal data was first put on firm
formal foundations in~\citet{robins1986new}.
It has subsequently
blossomed into a rich body of work; see~\citet{hernan2020causal} 
for a recent overview. Within this line of work, marginal structural models (MSMs)~\citep{robins2000marginal, hernan2001marginal} are commonly used to precisely define the causal quantity of interest, and are readily fit using IPW estimators. 
An MSM directly models the effect of the treatment $A_t$ on the outcome $Y_t$, without having to specify a conditional model for $Y_t$ given the treatment and covariates $(A_t, X_t)$).
Finding optimal estimators for causal quantities is also of interest, 
and much effort 
has gone into deriving semi- and non-parametric efficient and doubly robust estimators for this setting; see, e.g.,
\citet{Bang2005, van2018targeted, Lendle2017, Schomaker2019, luedtke2017sequential}.
In contrast, here we do not concentrate on MSMs but rather focus on modelling the outcome $Y_t$ parametrically and estimating causal effects using regression methods. This approach does result in
estimates of the marginal effect, but they arise from a conditional model that includes
both the observed covariates and treatment status rather than a marginal model only in terms of 
the treatment status.
Our motivations for taking this route is that regression methods are widely available and familiar to data analysts. 

In our setting, the causal effect is given by the g-formula (\cite{robins1986new}; see (\ref{eq:gform_full}))
and can be estimated by plugging in estimates for the unknown quantities.
When plugging in
parametric estimates, we will call such methods
\emph{parametric g-formula} methods; see~\citet{naimi2016} and \citet{vansteelandt2016revisiting} for recent
overviews and~\citet{mcgrath2020} for a software library implementing these methods.
Parametric g-formula methods have been scrutinized for
their vulnerability to the \emph{null paradox}~\citep{robins1997nullparadox}:
they can give inconsistent estimates of causal effects, especially
when there is no true effect.
\citet{McGrath2022} evaluate 
the degree of possible bias and provide recommendations, and~\citet{evans2021parameterizing} 
derive a set of models that avoid the null paradox. Here, we will also introduce a set of parametric g-formula methods that avoid the null paradox, although, unlike~\citet{evans2021parameterizing}, we are motivated
by regression modeling and introduce a distinct set of models.
In principle, one can plugin nonparametric estimates which then avoid the null paradox.
But, as the number of time points increases, we face an explosion of dimensionality
and so, once again, the estimate is inconsistent due to the curse of dimensionality.

In the present work, we consider several families of regression models: linear regression, log-linear regression, probit regression, and Cox regression.
We will focus on the latter to highlight our method's applicability to
challenging, nonlinear regression tasks.
In more detail, we pay special attention to the Cox proportional hazards model
with time-varying covariates~\citep{Tian2005, Wang2017}.
Correctly analyzing time-varying treatment effects with these
models does require the additional causal machinery mentioned above:
marginal structural Cox models have been developed
to allow for flexible specification ~\citep{Xiao2014} and
efficient estimation~\citep{Zheng2016,diaz2019statistical} of causal effects.
Similarly, parametric g-formula methods for survival data have been 
developed previously in~\citet{Keil2014parametric, young2014simulation, wen2021parametricsurvival}. 

Our approach is a \emph{regression with residuals} method~\cite{almirall2010structural, almirall2014time, wodtke2017estimating, wodtke2020regression}, which yield estimators for various causal estimands by running regression models on modified variables that residualize out the past. 
Regression-with-residuals methods were originally developed to estimate parameters in a structural nested mean model. These parameters sometimes determine
MSM parameters, however, and some recent work focuses explicitly on estimating
MSM parameters~\cite{wodtke2020regressionmarginal}.
In the context of the regression with residuals family of methods, the novelty of this work is that we consider several nonlinear outcome models (such as the Cox model) and show their relationship with MSM parameters. 

Our approach in the linear model case is closely related
to the rich body of work on identifiability and estimation in linear structural equation models.
In particular, in this setting our estimator is the usual maximum likelihood estimator (MLE) for Gaussian
linear models. However, we use a new parameterization that enforces certain orthogonality conditions
to yield simple computations. Leveraging orthogonality to compute 
the MLE in linear models goes back classically to~\citet{Frisch1933}, and ideas
similar to those in our work are implicit in the computational schemes in
\citet{Drton2009} and \citet{Nandy2017}; see also
\citet{shpitser2018acyclic}. 
Recently, \cite{guo2021efficient} showed that such estimators are
not only optimal for the Gaussian case, but are also
optimal among estimators that use only the empirical covariance matrix 
for (non-Gaussian) linear models.
More broadly, ideas of orthogonality are ever-present in 
econometrics and causal inference; see, e.g.,
\citet{Lancaster2002, kennedy2016semiparametric,Chernozhukov2018double}.
However, the role of orthogonality in the regression-with-residuals methods is distinct from these instances---
here, we use orthogonality to break certain treatment to covariate edges in the causal graph,
thereby removing causal paths that would otherwise invalidate the causal effect estimates derived from regression modeling.


\subsection{Our contribution}
We propose a new estimator, \emph{nonlinear regression with residuals}, to obtain causal estimates via regression adjustment for data with time-varying treatments and covariates. We develop the case of linear regression, log-linear regression, probit regression and Cox regression, and we expect that further cases can also be handled.
The method is computationally and conceptual simple; it requires only a sequence of standard regression fits and the operations are easily understood in terms of collinearity, a familiar topic in regression modeling.

Our presentation proceeds as follows. In Section~\ref{sec:setting} we introduce our notation, explain the difficulty with regression modeling in our setting, and introduce our estimator. In Section~\ref{sec:methods}, we show that the estimator is indeed estimating the causal effect for linear, log-linear, probit, and Cox regression. In Section~\ref{sec:experiments}, we evaluate the proposed method in simple simulated examples, and then in Section~\ref{sec:example} we demonstrate it's use in a realistic setting.

\section{What is the problem?}
\label{sec:setting}

\subsection{Notation}
Suppose we observe
data on $n$ individuals
over time.
The time ordered data on the $i^{\rm th}$ subject are
$$
\left(X_{1}^\i, A_{1}^\i,\right), \ldots, \left(X_{t}^\i, A_{t}^\i\right),\ldots, \left(X_{T}^\i, A_{T}^\i\right), Y^\i.
$$
Here,
$X_{t}^\i$ is a vector of covariates
$A_{t}^\i$ is treatment or exposure,
and $Y^\i$ is the outcome of interest. 
The vector $X_{t}^\i$ can include past values
of the outcome.
The reader should keep in mind the case where $X_t^\i$ 
corresponds to the same covariate measured at different time
points as the representative example of our setup. 
But this is not a requirement:
$X_t^\i$ could also have different dimensions
and represent different covariates for different $t$.
In what follows, we omit the superscript $(i)$
when considering a generic subject 
and employ the notation
$\overline{X}_t = (X_1,\ldots,X_t)$,
$\overline{X} = (X_1,\ldots,X_T)$,
$\overline{A}_t = (A_1,\ldots,A_t)$
and $\overline{A} = (A_1,\ldots,A_T)$.

We are interested in the effect of treatment on $Y$.
Let $Y^{\overline{a}}$
be the value that $Y$ 
would have
if the treatment variable
$\overline{A} = (A_1,\ldots,A_T)$
was set to a particular value
$\overline{a} = (a_1,\ldots,a_T)$.
We are interested in
\begin{equation} \label{eq:psi_def}
\psi(\overline{a}) = \E[Y^{\overline{a}}].
\end{equation}
This is the average value of $Y$ if we were to 
intervene to set $\overline A = \overline{a}$ for everyone in the 
population---it answers the question ``if we changed 
$\overline{a}$ in the same way for every unit, what would 
happen to the population?''
Our modeling choices will imply that $\psi$ 
is parameterized by a finite vector, 
denoted by $\theta \in \mathbb{R}^d$.
A basic question of interest is whether or not there is
any treatment effect, that is, whether or not $\psi$ is a constant.
A more precise question is exactly how the different coordinates of $\bar{a}$
affect the outcome. This corresponds to estimating the whole
function $\psi$.

\subsection{A linear regression example}
\label{subsec:phantoms_linear_regression_teaser}

To understand the shortcomings of standard regression modeling, we return to the motivating example on the left of Figure~\ref{fig:teaser}, described by the DAG in the right side of Figure~\ref{fig::dag}.
To facilitate explicit formulae for the effects seen in Figure~\ref{fig::dag}, we further assume $(A_1, X_2, A_2, Y, U) \sim \mathcal{N}(0, \Sigma)$, with covariance matrix $\Sigma$ determined by size of the effects $A_1 \to X_2$, $X_2 \to A_2$, $U \to X_2$ and $U \to Y$. We standardize all variables to have variance 1, and refer to these correlations as $\rho_{A\to X}, \rho_{X \to A}, \rho_{U \to X}$ and $\rho_{U \to Y}$, respectively. It suffices to vary $\rho_{U \to X}$ and set $\rho_{U \to Y} = 1$, which reduces the model to three parameters. 
The implied covariance matrix for the observed
quantities $(A_1, X_2, A_2, Y)$ is a function of
these three parameters:
\begin{equation*}
\begin{bmatrix}
1 & \rho_{A \to X} & \rho_{A \to X} \rho_{X \to A} & 0  \\
\rho_{A \to X} & 1 & \rho_{X \to A} & \rho_{U \to X}  \\
\rho_{A \to X} \rho_{X \to A} & \rho_{X \to A} & 1 &  \rho_{X \to A} \rho_{U \to X}  \\
0 & \rho_{U \to X} & \rho_{X \to A} \rho_{U \to X} & 1
\end{bmatrix}.
\end{equation*}
We assume that
the correlations are such that 
this is a valid covariance matrix. 
A 
calculation yields the population regression coefficients of $Y$ on $(A_1, X_2, A_2)$:
\begin{equation}
\label{eq:lin_reg_coeff}
\beta = \left(\frac{-\rho_{A \to X} \rho_{U \to X}}{1 - \rho_{A \to X}^2}, \frac{\rho_{U \to X}}{1 - \rho_{A \to X}^2}, 0\right)
\end{equation}
from which we confirm that the regression coefficient for $A_1$ is misleading: it is nonzero, even though there is no treatment effect. This is the result of the collider bias from including $X_2$ in the regression model. On the other hand, the regression coefficient for $A_2$ is correct in this case; including $X_2$ in the regression model correctly accounts for the confounding effect of $X_2$ on $A_2$. If we instead removed $X_2$ from the regression model, we would avoid the collider bias, but fail to account for the confounding, so the coefficient for $A_2$ would be biased. As previewed in our experiment, we will soon introduce an alternative to this regression model that correctly identifies the causal effects.

Notice that~\eqref{eq:lin_reg_coeff} provides a hint about when the regression model will fail badly. As $|\rho_{A \to X}|$ increases, the regression coefficient for $A_1$ deviates farther from zero. The magnitude of the collider induced bias varies as with impact of the unobserved variable times the magnitude of the effect from $A_1$ to $X_2$, $\rho_{A \to X}$. \textit{Thus, the collider bias can be of magnitude comparable to the effect of confounding}. This is an appreciable effect that cannot typically be ignored. In summary, the regression model has worse bias when there is a strong effect from the time-varying treatment to the time-varying covariate.

\subsection{Phantoms variables: why naive modeling fails}

\begin{figure}
\begin{center}
\begin{tikzpicture}[->, shorten >=2pt,>=stealth, 
node distance=1cm, noname/.style={ ellipse, minimum width=5em, minimum height=3em, draw } ]
\node[] (1) {$A_1$};
\node (2) [right= of 1] {$X_2$};
\node (3) [right= of 2] {$A_2$};
\node (4) [right= of 3] {$Y$};
\node[circle,fill=red!20,draw] (5) [below= of 3] {$U$};
\path (1) edge [thick] node {} (2);
\path (1) edge [thick, bend left=50pt] node {} (3);
\path (1) edge [thick, bend left=60pt] node {} (4);
\path (2) edge [thick] node {} (3);
\path (3) edge [thick] node {} (4);
\path (5) edge [thick] node {} (2);
\path (5) edge [thick]  node {} (4);
\end{tikzpicture}\hspace{1cm}
\begin{tikzpicture}[->, shorten >=2pt,>=stealth, 
node distance=1cm, noname/.style={ ellipse, minimum width=5em, minimum height=3em, draw } ]
\node[] (1) {$A_1$};
\node (2) [right= of 1] {$X_2$};
\node (3) [right= of 2] {$A_2$};
\node (4) [right= of 3] {$Y$};
\node[circle,fill=red!20,draw] (5) [below= of 3] {$U$};
\path (1) edge [thick] node {} (2);
\path (2) edge [thick] node {} (3);
\path (5) edge [thick] node {} (2);
\path (5) edge [thick]  node {} (4);
\end{tikzpicture}
\end{center}
\caption{\em  
Example from Robins and Wasserman (1997).
Treatments $(A_1,A_2)$, covariate $X_2$, outcome $Y$ and
phantom $U$.
(Left) The treatment affects the outcome $Y$. There are no unmeasured confounders.
(Right)
There is no treatment effect as there are no arrows from
$A_1$ to $Y$ or from 
$A_2$ to $Y$.
The covariate $X_2$ is a collider, meaning that two arrowheads meet at $X_2$.
This implies that $Y$ and $(A_1,A_2)$  are dependent conditional on $X_2$.}
\label{fig::dag}
\end{figure}
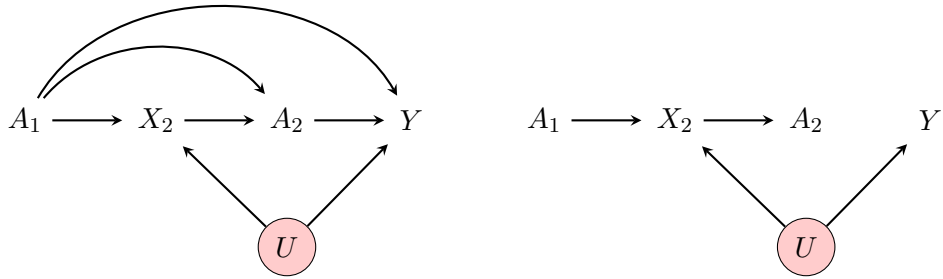

Why do straightforward regression models fail with time-varying treatments?
The heart of the issue is what we call \emph{phantom} variables: 
unmeasured variables that link measured variables with the response. 
For example, consider the examples
from~\citet{robins1997nullparadox} with measured variables
$(A_1,X_2,A_2,Y)$ and graphical structures in Figure~\ref{fig::dag}.
Here, the unmeasured $U$ is a phantom, linking the observed variables
$X_2$ with the response $Y$.
An example of a phantom is undelying health status of a subject. 
Then $U$ is likely associated with the covariates $X_2$ and outcome $Y$.
Note that, even in randomized studies, there will likely be
many phantoms. Phantoms are \emph{not} confounders. Unlike the case with
unmeasured confounders, even with unmeasured phantoms the causal 
effect is identified and can be estimated using appropriate methods.

However, phantom variables invalidate a naive regression analysis in the following way. 
The variable $X_2$ in Figure~\ref{fig::dag} is an observed confounder that must be included
in the model to get a consistent estimate of the treatment effect.
However, by (naively) including $X_2$, 
the analyst is conditioning on a post-treatment variable, 
which creates \emph{collider bias} and leads to
invalid estimates of, and inference for, the causal effect. 

For readers familiar with directed graphs,
we see from Figure \ref{fig::dag} that
$X_2$ is a collider and this is why $Y$ is conditionally dependent on
$(A_1,A_2)$ given $X_2$, even when there is no treatment effect.
For those not familiar with directed graphs,
recall that we computed the conditional correlation algebraically
in the previous section and saw that $Y$ and $A_1$ were not 
conditionally independent.

\subsection{Causal identifiability and the g-formula} \label{eq::g-methods}

Naive regression modeling fails
not because estimation
is impossible in this setting; the causal effect is
indeed identifiable, and so it can be estimated with the right methods.
\citet{robins1986new} showed that,
if there are no unmeasured confounders
(and subject to two technical conditions, no-interference and positivity, described in Appendix~\ref{app:identification}) the causal effect
$\psi(\overline{a})$ in (\ref{eq:psi_def}) is given by the \textit{g-formula}:
\begin{equation}
\label{eq:gform_full}
\psi(\overline{a}) = 
\int\cdots\int 
\mu(\overline{a},\overline{x})\ 
\prod_{t=1}^T p( x_t |  \overline{x}_{t-1},\overline{a}_{t-1}) d x_t
\end{equation}
where
$$
\mu(\overline{a},\overline{x}) =
\E[ Y| \overline{X} = \overline{x},\overline{A} = \overline{a}].
$$
The intuition for the formula is this as follows.
Take the joint density
$$p(x_1)p(a_1|x_1)p(x_2|x_1,a_1)p(a_2|x_1,x_2,a_1)\cdots.$$
Now replace each
$p(a_t | {\rm past})$ with a point mass at $a_t$ (as if we 
intervened to set the treatment to be $a_t$ for all units)
and take the mean of $Y$ from the resulting distribution.

Using the g-formula as the starting point
for estimation has led to the development of
\emph{g-formula methods}.
Nonparametric approaches  
estimate the density of $p(x_s \mid \bar{x}_{s-1}, \bar{a}_s)$,
but this has high statistical complexity: many samples are needed for reliable density estimation.
On the other hand, parametric models are known
to be vulnerable to the \emph{null paradox}: 
naively plugging in parametric models can result in 
inconsistent estimates even under the null~\citep{robins1997nullparadox}. 
In this work, we show how to use regression models that 
fit well with the g-formula and avoid the null paradox. 
This enables the use of familiar regression tools
to correctly estimate causal effects.

\subsection{Proposed method: nonlinear regression with residuals}

To adapt regression models for use with time-varying
treatments and covariates, we propose 
\emph{nonlinear regression with residuals}. In particular,
we proposed modifications to Poisson regression,
Cox regression and so on, that correctly handle time-varying
treatments and covariates. 
The key idea is that before model fitting, we replace each $X_t$
with an \emph{orthogonalized} version $\Xt_t$, where
$\Xt_t$ is the residual of $X_t$ when we regress out all 
past variables $(X_1, A_1, \dots, X_{t-1}, A_{t-1})$, as 
outlined in Algorithm~\ref{alg:resid_regression}.
To obtain causal estimates, we then regress $Y$
onto $(\Xt_1,A_1,\dots,\Xt_T, A_T)$ and extract the fitted coefficients for $A_1,\dots,A_T$.

The motivation is that
because naive regression modeling fails due to
collider bias, if we can remove the collider bias we can obtain
reliable estimates of the causal effect. 
To remove the colider bias,
we remove the edges from $(\overline{X}_{t-1}, \overline{A}_{t-1})$ to $X_t$
by residualizing out the past information. For example, in Figure 3,
the orthogonalization of $X_2$ removes the edge from $A_1$ to $X_2$, so that 
$X_2$ is no longer a collider.
In the language of regression modeling, the confounding
is due to collinearity between treatment variables and subsequent 
covariates, so nonlinear regression with residuals removes this collinearity 
before fitting the final regression model.

The remainder of this work will make precise the causal guarantees
of this method in various settings.

\begin{algorithm}
\caption{\em Nonlinear regression with residuals}\label{alg:resid_regression}
\begin{algorithmic}
\State \textbf{Input:} data \ $\left(\overline{X}^\i, \overline{A}^\i, Y^\i\right) \text{ for subjects } i=1,\dots,n$, regression fitting subroutines $\regression^{(X)}$ and $\regression^{(Y)}$
\vspace{.2cm}
\For{t = 1,\dots,T}
	\State $\hat{g}_t = \regression^{(X)}\left(X_t \sim (\overline{X}_{t-1}, \overline{A}_{t-1})\right)$ 
	\Comment{fit regression model with observed data}
	\State $\Xt_t^\i = X_t^\i - \hat{g}_t \left(\overline{X}_{t-1}^\i, \overline{A}_{t-1}^\i \right)$ for $i=1,\dots,n$
	\Comment{compute residuals}
\EndFor
\vspace{.2cm}
\State $\hat{f} = \regression^{(Y)}\left(Y \sim (\Xt_1,A_1,\dots,\Xt_T,A_T)\right)$
\Comment{nonlinear regression with residuals fit}
\vspace{.2cm}
\State \textbf{Output: } $\hat{f}$  \Comment{$A_t$ coefficients in the orthogonalized model are causal estimates}
\end{algorithmic}
\end{algorithm}

\section{The causal estimand of nonlinear regression with residuals}
\label{sec:methods}

The central result of this work is that regression coefficients from 
Algorithm~\ref{alg:resid_regression} are valid estimates of
causal effects.
In this section, we make this precise for a variety of 
cases: linear models, log-linear models, probit regression 
and Cox proportional hazard model. In each case,
given certain conditions on the time-varying covariates $\overline{X}$, 
we show that the regression coefficients from
Algorithm~\ref{alg:resid_regression} correspond
to the plug-in estimator from the g-formula in~\eqref{eq:gform_full}. 
This implies that the estimates are consistent for
a causal parameter. We will now make explicit for each case.

\subsection{ Warm-up: regression with residuals  in the Gaussian linear model}

As an initial simple example, 
we return to the linear structural causal model from Section~\ref{subsec:phantoms_linear_regression_teaser}, treating
this case in detail. 
However, we will now only assume that the variables are jointly Gaussian with
graph given in the left side of Figure~\ref{fig::dag}. 
Our goal is to estimate
the total causal effect \eqref{eq:psi_def} when setting the treatment $\bar{A}$ to some value $\bar{a}$.
Since our main goal is to show that our estimate from Algorithm~\ref{alg:resid_regression} 
is consistent for the true causal effect,
for now we will work in the population (infinite data) setting.

We seek to estimate the causal effect 
of $(A_1, A_2)$ on $Y$, which is given by the g-formula in~\eqref{eq:gform_full}
with
$$
\mu(a_1, x_2, a_2) =
\E[ Y| A_1 = a_1, X_2 = x_2, A_2 = a_2] = \beta^\naive_1 a_1 +  \lambda^\naive_2 x_2 + \beta^\naive_2 a_2,
$$
where
\begin{equation}
\label{eq:gamma_def}
\left(\beta^\naive_1, \lambda^\naive_2, \beta^\naive_2\right) = \argmin_{(\beta_1, \lambda_2, \beta_2)}
\E\left(Y-\beta_1 A_1 - \lambda_2 X_2 - \beta_2 A_2 \right)^2
\end{equation}
are the population regression 
coefficients of $Y$ onto $(A_1, X_2, A_2)$.
Plugging this into the g-formula in~\eqref{eq:gform_full} and simplifying yields
\begin{equation*}
\psi(a_1, a_2) = \underbrace{\left(\beta^\naive_1 + \frac{\cov(A_1, X_2)}{\var(A_1)} \lambda^\naive_2\right)}_{\theta_1} a_1 + \underbrace{\beta^\naive_2}_{\theta_2} a_2.
\end{equation*}
This calculation shows that
$\psi(a_1, a_2)$ is linear in $(a_1, a_2)$, with coefficients $(\theta_1, \theta_2)$ as
the causal effects of $A_1$ and
$A_2$ on $Y$, respectively. Because $\theta_1 \neq \beta^\naive_1$, this calculation also shows that 
the naive regression coefficient of $A_1$ is not consistent for the causal effect of $A_1$.
(In this example, naive regression does yield a consistent estimate of the
causal effect of $A_2$; this is because there is no collider path from $A_2$ to $Y$.)


Next, we show that nonlinear regression with residuals 
in Algorithm~\ref{alg:resid_regression} yields consistent estimates of $(\theta_1, \theta_2)$.
The method consists of regressing $Y$ on $A_1$, $A_2$ and the \emph{orthogonalized}
version of $X_2$:
\begin{equation}
\label{eq:xtilde_def}
\Xt_2 := X_2 - \frac{\cov(A_1,X_2)}{\var(A_1)} A_1,
\end{equation}
yielding regression coefficient estimates for $(A_1, A_2)$:
\begin{equation}
\label{eq:ortho_msm_fit}
({\beta}_1^\adj, {\beta}^\adj_2) =\argmin_{\beta_1, \beta_2} \min_{\lambda}
\E \left(Y - \beta_1 A_1 - \beta_2 A_2 - \lambda \Xt_2\right)^2.
\end{equation}
The intuition is that it is safe to include $\Xt_2$ in
the regression model because it does not include any collider bias: $\Xt_2$ is
independent of $A_1$.
Our key result is that this is a consistent estimator for
the causal effect:

\begin{proposition}[Nonlinear regression with residuals for Gaussian linear models]
\label{prop:ortho_msm_fit}
Assume the no unmeasured confounding, no interference, and positivity conditions (Appendix~\ref{app:identification}).
In the setting above, $\psi$ is linear in $(a_1, a_2)$:
\begin{equation*}
\psi(a_1, a_2) = \theta_1 a_1 + \theta_2 a_2.
\end{equation*}
Moreover, the nonlinear regression with residuals coefficients
correspond to the causal parameters:
\begin{equation*}
({\beta}_1^\adj, {\beta}_2^\adj) = (\theta_1, \theta_2).
\end{equation*}
\end{proposition}

We include the proof of the preceding proposition because it is straightforward and illustrates the key idea.

\begin{proof}
Starting with~\eqref{eq:ortho_msm_fit} 
and the definition of $\Xt_2$, we have
\begin{equation}
\label{eq:ortho_minimizer_linear}
({\beta}_1^\adj, {\beta}^\adj_2) =\argmin_{\beta_1, \beta_2} \min_{\lambda}
\E \left(Y - \beta_1 A_1 - \beta_2 A_2 - \left[X_2 - \frac{\var(A_1)}{\cov(A_1,X_2)} A_1 \right]\lambda\right)^2.
\end{equation}
The optimization problems in~\eqref{eq:ortho_minimizer_linear} and~\eqref{eq:gamma_def} are reparameterizations of each 
other so they take the same minimum value. We will thus obtain a set of parameters for (\ref{eq:ortho_minimizer_linear}) 
that achieves the optimal value in~\eqref{eq:gamma_def}, which proves
that they are also minimizers of~\eqref{eq:ortho_minimizer_linear}.

First set
${\beta}^\adj_2 = \beta^\naive_3 = \theta_2$ and rewrite
\begin{align*}
{\beta}_1^\adj =\argmin_{\beta_1} \min_{\lambda}
\E \left(Y - \left[\beta_1 - \frac{\var(A_1)}{\cov(A_1,X_2)} \lambda\right] A_1 - \theta_2 A_2 - X_2 \lambda\right)^2,
\end{align*}
from which we see that the minimizer for $\lambda$ above 
is $\lambda^\naive_2$, and thus
\begin{equation*}
{\beta}^\adj_1 - \frac{\var(A_1)}{\cov(A_1,X_2)} \lambda^\naive_2 = \beta^\naive_1.
\end{equation*}
\end{proof}



\subsection{Linear model without Gaussianity}
\label{subsec:linearity}

We next generalize the estimator from the previous 
to the case without the Gaussian assumption, and
consider an arbitrary number of time steps $T$.

Let $\Xt = (\Xt_1,\dots,\Xt_T)$ be the orthogonalized covariates where
\begin{equation}
\label{eq:xtilde_def_nonlinear}
\Xt_t := X_t- \E[X_t \mid \overline{X}_{t-1}, \overline{A}_{t-1}].
\end{equation}
From here, we assume that the mean of $Y$
is linear in the treatment and orthogonalized covariates:
\begin{equation}
\label{eq:linearity_assumption}
\E[Y \mid \overline{X}, \overline{A}] = \alpha + \beta^\top \overline{A} + \lambda^\top \Xt. 
\end{equation}
We emphasize that we are not assuming anything about the
distribution of $(\overline{X}, \overline{A})$ except that the 
conditional mean of $X_t$ given the past exists.

In the setting above, it is again the case that 
a regression fit of $Y$ onto
$(\Xt, \overline{A})$ corresponds to a g-formula 
plug-in estimator.
Specifically, we consider the nonlinear regression with residuals fit
\begin{equation}
\label{eq:beta_ortho_linear}
\beta^\adj = \argmin_{\beta} \min_{\alpha,\lambda} \E\left(Y - \alpha - \beta^\top \overline{A} - \lambda^\top \Xt \right)^2.
\end{equation}
These are causal estimates, as stated formally next.
\begin{proposition}[Nonlinear regression with residuals for linear models]
\label{prop:ortho_linear_general}
Assume the no unmeasured confounding, no interference, and positivity conditions (Appendix~\ref{app:identification}).
In the setting above, $\psi(\overline{a})$ is linear:
\begin{equation*}
\psi(\overline{a}) = \theta_0 + \theta^\top \overline{a},
\end{equation*}
and the nonlinear regression with residuals coefficients in~\eqref{eq:beta_ortho_linear}
are the causal parameters: $\beta^\adj = \theta$. 
\end{proposition}

Notice again that this is a population level result. 
In practice, we do not observe $\Xt$, so Algorithm~\ref{alg:resid_regression}
instead estimates the distribution of 
$X_t \mid \overline{X}_{t-1}, \overline{A}_{t-1}$ 
and uses this to obtain an approximation of $\Xt_t$. 

\begin{remark}
These results for the linear case are known in the literature---see Section~\ref{sec:relatedwork}.
\end{remark}

\subsection{Modeling with orthogonalized variables}
We pause here to discuss the interpretation of the outcome model assumed in~\eqref{eq:linearity_assumption}, where the mean of $Y$ is assumed 
to be linear in the variables $\Xt_t$ and $A_t$ for $t = 1,\dots,T$. How
does this relate to the familiar assumption that $Y$
is linear in $X_t$ and $A_t$? In fact, these two outcome models are the same in the most typical case: when $X_t$ is modeled as depending linearly on $(\overline{X}_{t-1}, \overline{A}_{t-1})$. That is, the assumption in~\eqref{eq:linearity_assumption} is
then equivalent to
\begin{equation*}
\E[Y \mid \overline{X}, \overline{A}] = \alpha + \beta^\top \overline{A} + \lambda^\top \overline{X}. 
\end{equation*}
Although cosmetically different, our outcome model for orthogonality regression corresponds to a routine setup.
This continues to be the case in the nonlinear regression models that we turn to next.

\subsection{Log-linear model}

As a first extension of regression with residuals to nonlinear models, we 
consider log-linear models. 
As before, we define the orthogonalized covariate $\Xt$ as 
in~\eqref{eq:xtilde_def_nonlinear}. Additionally, we assume 
that it follows a location model: $X_t \stackrel{d}{=} g_t(\bar{a}_{t-1}, \bar{X}_{t-1}) + \epsilon_t$
for some function $g_t$ and independent random variable $\epsilon_t$.
Lastly, we assume the log of the mean of $Y$ 
is linear in the treatment and orthogonalized covariates:
\begin{equation}
\label{eq:log-linearity_assumption}
\E[Y \mid \Xt, \overline{A}] = \exp\{\alpha + \beta^\top \overline{A} + \lambda^\top \Xt\}. 
\end{equation}

In this setting, regression coefficients from the log-linear regression in~\eqref{eq:log-linearity_assumption}
are causal estimates, as stated next.
\begin{proposition}[Nonlinear regression with residuals for log-linear models]
\label{prop:ortho_loglinear}
Assume the no unmeasured confounding, no interference, and positivity conditions (Appendix~\ref{app:identification}).
In the setting above, $\psi$ is log-linear:
\begin{equation*}
\psi(\overline{a}) = \exp\{\alpha' + \theta^\top \overline{a}\},
\end{equation*}
for some $\alpha' \in \mathbb{R}$.
Moreover, $\theta$ is equal to $\beta$
in the full log-linear model in~\eqref{eq:log-linearity_assumption}.
\end{proposition}
That is,
as with linear models, the g-formula simplifies
such that the causal parameters coincide with the 
regression coefficients.

\subsection{Gaussian-probit model}

As a second foray into nonlinear models, we consider probit
models for a binary response $Y$.
As before, the nonlinear regression with residuals coefficients from Algorithm~\ref{alg:resid_regression} 
will be valid causal estimates, although the story
is slightly more complex in this model, which will
be explained soon.
As before, we define the orthogonalized covariate $\Xt$ as 
in~\eqref{eq:xtilde_def_nonlinear}. We now additionally assume 
that that the covariates follow a Gaussian location model:
$X_t \stackrel{d}{=} g_t(\bar{a}_{t-1}, \bar{X}_{t-1}) + \epsilon_t$
for some function $g_t$ and independent random variable 
$\epsilon_t \sim \mathcal{N}(0, \sigma_t^2)$.
Lastly, we assume that the binary response $Y$ 
follows a probit model:
\begin{equation}
\label{eq:probit_assumption}
P\left(Y = 1 \mid \overline{A}, \Xt \right) = \Phi\left(\alpha + \beta^\top \overline{A} + \lambda^\top \Xt \right),
\end{equation}
where $\Phi$ is the CDF of the standard normal distribution. 
Similar to before, the g-formula evaluates to a probit
model on $\bar{A}$, but now with re-scaled coefficients, as stated next.
\begin{proposition}[Nonlinear regression with residuals for probit-Gaussian models]
\label{prop:ortho_probit}
Assume the no unmeasured confounding, no interference, and positivity conditions (Appendix~\ref{app:identification}).
In the setting above, $\psi$ follows a probit model:
\begin{equation*}
\psi(\overline{a}) = \Phi(\alpha' + \theta^\top \overline{a}).
\end{equation*}
Moreover, $\theta = \beta / \sqrt{1 + \sigma^2}$, where 
$\beta$ are the regression coefficients 
from~\eqref{eq:log-linearity_assumption} and
$\sigma^2 = \lambda_1^2\sigma_1^2 + \dots + \lambda_T^2\sigma_T^2$ 
with $\lambda$ from~\eqref{eq:probit_assumption}.
\end{proposition}

\paragraph{Why is there re-scaling?}

One may wonder why the probit regression coefficients 
need to be re-scaled to recover the $g$-formula. The intuition 
is that with a probit regression (or a logistic regression)
the magnitude of the coefficient vector is encoding the
certainty/uncertainty. Larger magnitude corresponds to 
higher confidence. In the marginal model, we have less
certainty, so the magnitude of the coefficients shrink.
The direction of the coefficient vector remains unchanged,
which means that a coefficient in the marignal model is zero
only when the corresponding coefficient in the full model
is zero.

\paragraph{Beyond the probit model}

The calculation in this seemingly-specific model class
is more general than it first appears, 
in that the results will approximately hold even when the 
regression in~\eqref{eq:probit_assumption}
is not a probit regression, and when the residuals
are not exactly Gaussian. The reason is the
central limit theorem; with many time points,
we can expect $\sum_t \lambda_t \Xt_t$ to be approximately
Gaussian. As an example, consider logistic regression. 
Logistic regression is a \emph{threshold model}:
it can be thought of as the probability of a logistic
random variable falling below a threshold: 
\begin{equation*}
P(Y = 1 \mid \overline{A}, \overline{X}) = P(Z \le \overline{A} \beta + \overline{X} \lambda'),
\end{equation*}
where $Z$ is a logistic random variable. After orthogonalization
and an application of the g-formula, we obtain
\begin{equation*}
\psi(\bar{a}) = P(Z - \sum_t \lambda_t \Xt_t \le \overline{a} \beta) \approx \Phi(\overline{a} \beta / \sigma^2),
\end{equation*}
where $\sigma^2$ is the variance of $Z - \sum_t \lambda_t \Xt_t$. 
Thus, this marginalized logistic model approaches
a probit model as the variance of 
$\sum_t \lambda_t \Xt_t$ grows large.
Roughly speaking, threshold models like logistic regression 
converge to probit models when marginalizing out
many time-varying covariates.

\subsection{Cox proportional hazards model}

Lastly, we consider the Cox propotional hazards model~\citep{Cox1972},
the most commonly used model for modeling survival times.
As before, we define the orthogonalized covariate $\Xt$ as 
in~\eqref{eq:xtilde_def_nonlinear}. We additionally assume 
that the the orthogonalized covariates follow a Gaussian location model.
That is, we are assuming 
$X_t \stackrel{d}{=} g_t(\bar{a}_{t-1}, \bar{X}_{t-1}) + \epsilon_t$
for some function $g_t$ and independent random variable 
$\epsilon_t \sim \mathcal{N}(0, \sigma_t^2)$.
Lastly, we assume that the survival time $Y$ follows a proportional
hazard model:
\begin{equation}
\label{eq:cox_model_assumption}
h(t \mid \overline{A}, \overline{X}) = h_0(t)\cdot \exp\{\beta^\top \bar{A} + \lambda^\top \Xt\},
\end{equation}
where $h_0$ is a unknown baseline hazard rate.
Recall that this implies that the survival probabilities
are 
\begin{equation*}
P(Y \ge t \mid \overline{A}, \overline{X}) = \exp\left\{-H_0(t) \exp\{\beta^\top \bar{A} + \lambda^\top \Xt\} \right\},
\end{equation*}
where
\begin{equation*}
H_0(t) = \int_0^t h_0(t')dt'
\end{equation*}
is the baseline cumulative hazard function.

We next apply the g-formula to show that the regression
coefficients from fitting the regression model above
have a causal meaning.
In particular, we will show that the implied g-formula 
has the form of
a Cox model with an additional random effect term,
and coefficients matching those from 
the regression in~\eqref{eq:cox_model_assumption}, as stated next.

\begin{proposition}[Orthogonalized Cox regression]
\label{prop:ortho_cox}
Assume the no unmeasured confounding, no interference, and positivity conditions (Appendix~\ref{app:identification}).
In the setting above, $\psi$ can be expressed as follows:
\begin{equation}
\label{eq:cox_marginal_model}
\psi(\bar{a}, t) := P(Y^{\bar{a}} \ge t) = \int \exp\left\{-H_0(t) \exp\{ \theta^\top \overline{A} + Z\} \right\} P(Z = z) dx,
\end{equation}
where $Z \sim \mathcal{N}(0, \sigma^2)$.
Moreover, the causal parameters $\theta$ are identical
to the coefficients $\beta$ from the
full model in~\eqref{eq:cox_model_assumption},
and $\sigma^2 = \sum_t \lambda_t^2 \sigma_t^2$.
\end{proposition}

The model arising from~\eqref{eq:cox_marginal_model} is a cousin 
of the usual Cox model; it is a Cox model with an additional random effects term
that is marginalized out. Put another way,
notice that this is the survival curve for a
population comprised of individuals with (Cox model) hazard functions
\begin{equation*}
h(t \mid \overline{A}, Z) = h_0(t) \exp\{\beta^\top \bar{A} + Z\},
\end{equation*}
where $Z \sim \mathcal{N}(0, \sigma^2)$ is a random effect varying across
each member of the population.
Thus, the regression from the full Cox model 
in~\eqref{eq:cox_model_assumption} has
a sensible causal interpretation: it is the causal
effect in a random-effects Cox regression model.
Note however, that this is not the same estimand 
as the usual marginal structural Cox model~\citep{hernan2000marginal}.

\subsection{Confidence intervals}

All of the models that we are fitting are regular, finite dimensional
parametric models.
The orthogonalized least squares estimators
are $M$-estimators.
It follows from Theorem 5.21 of 
\cite{van2000asymptotic}
that
the estimates are asymptotically normal, converging at rate $n^{-1/2}$.
The bootstrap provides valid asymptotic confidence intervals by Theorem 23.7 of 
\cite{van2000asymptotic}.

\section{Numerical experiments}
\label{sec:experiments}

We next investigate the regression with residuals estimator numerically for the linear model and the Cox regression model.
In both cases, we investigate two aspects of performance. First, we demonstrate
that they are valid: they produce estimated effects of zero (up to sampling variability)
under the null. Second, we compare the efficiency of the nonlinear regression with residuals estimator to 
the IPW estimator. Code reproducing these experiments is available at \url{https://github.com/stephenbates19/causal_orthogonalization}.

\subsection{Linear model}
\subsubsection{Validity under the null}
We consider the linear regression setting of Section~\ref{subsec:linearity} 
with $T=5$ time points. We simulate binary treatment assignments $A_t$ and
covariates $X_t$ whose mean depends on $A_{t-1}$.
Moreover, the treatment assignment $A_t$ depends on $X_{t}$,
and there is a phantom $U$, so there is possible collider bias.
Lastly, the response $Y$ is heavy-tailed and asymmetric.
There is no causal effect from $A_t$ on $Y$.
We treat the causal effect from $A_{t-1}$ to $X_t$ as a control 
parameter, investigating the behavior across a range of values.
In this experiment, we take a sample of 1 million data points,
so the results are approximately the results 
of the population regression.

The left panel of Figure~\ref{fig:linear_nongaussian_exp}, 
shows the results of a regression of $Y$ onto $(X_t, A_t)_{t = 1}^5$ (denoted \emph{naive})
with regression with regression---a regression of $Y$ onto $(\Xt_t, A_t)_{t = 1}^5$ 
(denoted \emph{ortho}).
In particular, we report the magnitude of the largest coefficient
of $A_1,\dots,A_5$ in each setting. Because there is no
causal effect, the ground-truth value is zero. However,
we observe that the naive regression produces large 
coefficients when there is a large effect size from
the treatment to the covariate. This is caused by collider bias, 
and it was anticipated by our analytic calculations 
in the Gaussian case. 
By contrast, nonlinear regression with residuals 
produces coefficients that are nearly zero 
(they are not exactly zero because we take the maximum of five numbers with Monte Carlo variability).
Thus, nonlinear regression with residuals is producing 
valid causal estimates in this case.

\subsubsection{Evaluating efficiency}
Continuing with this setting, we 
next investigate the efficiency of nonlinear regression with residuals
compared to fitting a MSM with IPW; see~\citet{vanderWal2011} for 
more details of the latter approach. 
We consider two cases.
In the first case, we have high overlap: $A_t$ is independent
of $X_{t-1}$, so the propensity score is always the same. In the
second case, we have poor overlap: $A_t$ is equal to the sign
of $X_{t-1}$ with probability 95\%. In this case, the IPW weights
are more extreme. Lastly, we vary the strength of the 
effect from $U$ to $X_t$ as a control parameter.

\begin{figure}
\begin{subfigure}[b]{.49\textwidth}
\centering
\includegraphics[height = 2.25in]{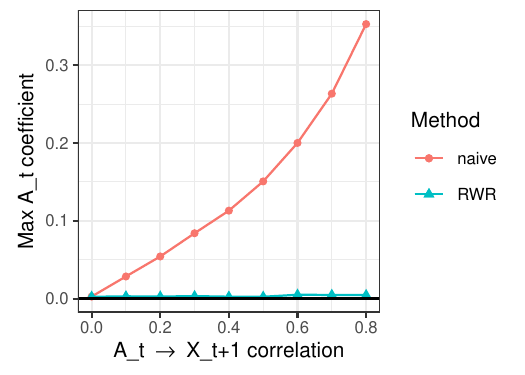}
\subcaption{Consistency}
\end{subfigure}
\hfill
\begin{subfigure}[b]{.49\textwidth}
\centering
\includegraphics[height = 2.25in]{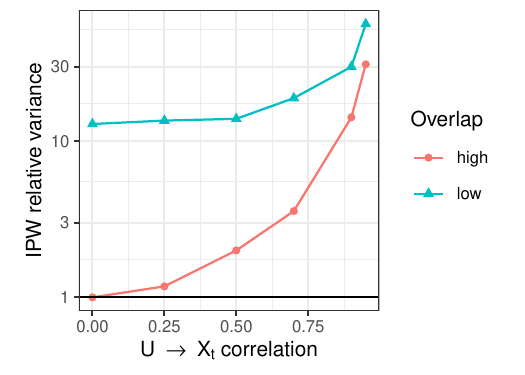}
\subcaption{Efficiency compared to IPW}
\end{subfigure}

\caption{\em Results of the linear, non-Gaussian simulation experiment. (a) Estimates of the causal effect with naive regression adjustment versus regression with residuals (RWR).
The true effect is zero.
(b) Relative variance of IPW compared to nonlinear regression with residuals in high and low overlap scenarios.}
\label{fig:linear_nongaussian_exp}
\end{figure}

We report the results in the right 
panel of Figure~\ref{fig:linear_nongaussian_exp}.
We find that as the effect from $U$ to $X_t$ increases 
in strength, nonlinear regression with residuals outperforms IPW by a larger margin.
This makes sense; nonlinear regression with residuals is able to regress out
the variation from $X_t$ (and by extension, part of $U$), but IPW is not
able to regress this out. Thus, for larger effects from $U$ to $X_t$,
nonlinear regression with residuals will have correspondingly more precise 
estimates. Secondly, the performance of IPW is worse when there is 
less overlap. This is a well-known problem with IPW; 
when there is poor overlap the weights become extreme, leading
to a small effective sample size and less efficiency. We emphasize
that the overlap in this case is not implausible; at each step any
treatment assignment can be seen with probability at least 5\%. Nonetheless,
IPW has poor relative efficiency.

\subsection{Cox proportional hazards model}

\subsubsection{Validity under the null}
We next demonstrate nonlinear regression with residuals with the Cox model with
a synthetic example across 5 time points.
The treatment assignment is binary, and is 
assigned based on the covariate at the 
previous time point.
The covariate is a location model based on the 
phantom, past treatments, and independent Gaussian noise.
The phantom in this case is the baseline hazard rate
for each subject, which spans a factor of 3 in the population.
(That is, the unhealthiest person has a hazard rate three
times larger than the healthiest person.)
We examine the null case where there is no treatment effect 
of $A_t$ on $Y$.
We treat the causal effect from $A_{t-1}$ to $X_t$ as a control 
parameter---larger values correspond to more collider bias.

We compare two approaches in this setting. First, we fit a Cox model 
with four covariates: $X_t$, $A_t$, the cumulative average of $X_t$, 
and the cumulative average of $A_t$, calling this approach \emph{naive}. 
We will report the coefficient of the cumulative average of $A_t$.
Since this a null case, these should be zero.
We compare this with a Cox model with covariates 
$X_t$, $A_t$, the cumulative average of $\Xt_t$, 
and the cumulative average of $A_t$, denoted \emph{RWR}.
In this case, $\Xt_t$ is computed by simply taking the residuals 
of a linear regression fit of $X_t$ on $A_{t-1}$.

\begin{figure}
\centering
\includegraphics[height = 2.25in]{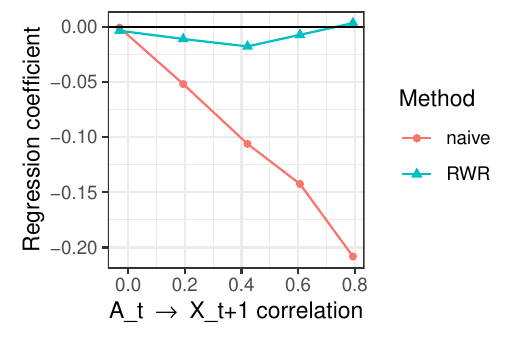}
\caption{\em Results of the Cox model experiment. We plot the regression coefficient corresponding to the cumulative average of past treatment assignments 
in the naive Cox regression model and in the regression-with-residuals Cox regression
model (RWR).}
\label{fig:cox_ortho}
\end{figure}

We report the results in Figure~\ref{fig:cox_ortho}, showing the average
of $250$ runs with $n=2,000$ subjects. We find that 
the naive regression model leads to nonzero coefficients, even under
the null. The bias gets worse as the size of the causal effect
from $A_{t-1}$ to $X_t$ increases, which is anticipated by our theoretical
results in the linear Gaussian model and previous simulations in the
(non-Gaussian) linear simulation. By contrast, the orthogonalized
regression yields an estimate near across all regimes.

\subsubsection{Evaluating power}
Next, we consider the same data generating mechanism as in the previous example, but now with a treatment effect.
In particular, $A_t = 1$ (i.e., receiving treatment) reduces the risk of an event during that next time step by constant factor. 
In this case, our goal is to detect if the treatment effect is present, so we will
check whether the confidence intervals for the coefficient for $A_t$ include zero. (Recall that the estimand for these two methods is different, but both are zero under the null. Therefore, checking power provides us a fair way to compare the efficiency.)
We consider IPW as implemented in~\citet{vanderWal2011} and nonlinear regression with residuals, but
but not the naive method, since we know it is not a valid estimator of the causal effect.
In each trial, we simulated $n=2,000$ individuals and record the resulting coefficient for $A_t$. 
We do this for many generated random data sets, and report the coefficients 
after standardizing them to have variance one (i.e., a z-score for the fitted coefficient).

\begin{figure}
\centering
\includegraphics[height= 2.5in]{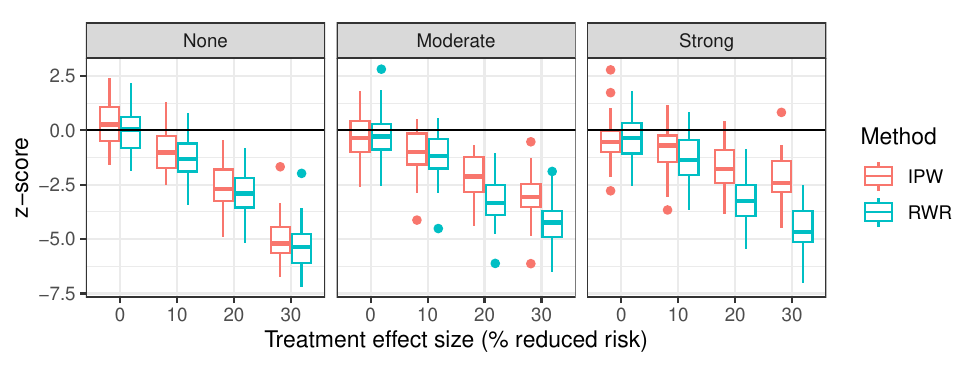}
\caption{\em The z-score for the causal effect with IPW and nonlinear regression with residuals (RWR). 
The three panels correspond to the strength of the effect from $X_t$ to $A_t$ (higher means less overlap).}
\label{fig:cox_zscore}
\end{figure}

We vary the strength of the treatment effect size and the strength of the effect of $X_t$ on $A_t$ 
as two control parameters. The first is measured as the fraction of reduced risk, and the
second varies across three levels, described precisely in the software accompanying this manuscript.
We report the results in Figure~\ref{fig:cox_zscore}. We find that as
the treatment effect increases, the z-scores separate away from zero
in the negative direction, as expected. Turning to the sensitivity to
the $X_t$ to $A_t$ effect size, we find that nonlinear regression with residuals is
mostly unaffected by this parameter. 
However, as this parameter grows, IPW becomes less effective at
detecting the treatment effect. This is expected; as the
effect grows, IPW produces more extreme weights and thus
has more variability.

\section{A realistic example}
\label{sec:example}

Lastly, we consider a synthetic observation HIV data set, introduced in the \texttt{ipw} package to evaluate IPW-based fitting methods. The goal is to evaluate whether a treatment regimen affects survival time. The data comprises 1200 individuals, measured for 15 time points, on average.  There are two baseline covariates, age and sex, as well as a time-varying covariate: the square root of the CD4 cell count, which decreases as HIV progresses. In this setting, patients initiate treatment during one of the measured time points, and then continue the treatment until the study concludes or a death occurs.

\begin{figure}
\begin{subfigure}[b]{.49\textwidth}
\centering
\includegraphics[height = 2in, clip, trim = 0 20 0 0]{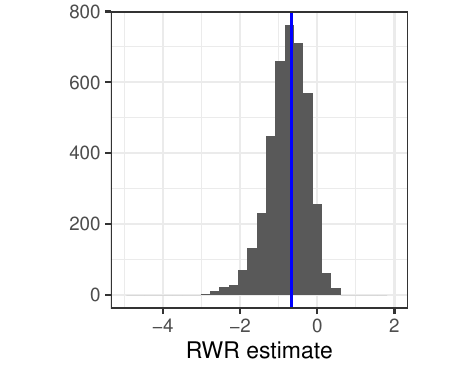}
\subcaption{Nonlinear regression with residuals}
\end{subfigure}
\begin{subfigure}[b]{.49\textwidth}
\centering
\includegraphics[height = 2in, clip, trim = 0 20 0 0]{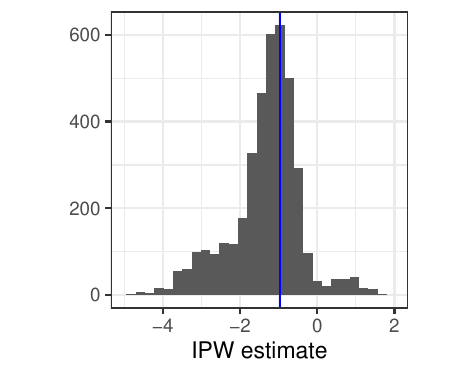}
\subcaption{IPW}
\end{subfigure}
\hfill

\caption{\em Estimates on bootstrap subsamples of the synthetic HIV survival data. The vertical blue line shows the point estimate from the full data.}
\label{fig:hiv_bootstrap_fig}
\end{figure}

We compare the estimate from the orthogonalized Cox regression estimator and the IPW estimator with a Cox marginal model. For nonlinear regression with residuals, we use Cox regression with the two baseline covariates, time-varying covariate, and time-varying treatment. For the MSM, we use a Cox regression model with the two baseline covariates and the time-varying treatment. (Recall MSMs never include the time-varying covariate.)
Both estimates suggest a protective effect, but the results are not statistically significant; using the bootstrap estimate of standard error the z-score for the orthogonalized estimate is -1.37 and that for the IPW estimate is -1.08. The basic bootstrap confidence intervals~\citep{davison1997boostrap} at the 90\% level are $(-1.32, .35)$ for the nonlinear regression with residuals estimator and $(-1.82, 1.21)$ for the IPW estimator. See Figure~\ref{fig:hiv_bootstrap_fig} for a visualization of the point estimates and bootstrap replicates. A careful look at the histogram of bootstrap values indicates one further point of concern about the IPW estimate; the bootstrap values are biased downwards, indicating possible positive bias to the IPW point estimate. 
Nonetheless, the IPW and nonlinear regression with residuals estimates are generally consistent with each other in this example, a good sign given the rather different modeling choices underlying the two methods. 

\section{Discussion}
\label{sec:discussion}

In this work, we have formulated a causal
estimator based on regression models with
time-varying treatments and time-varying covariates.
At its core, our method builds on the connection
between the causal concepts of collider bias 
and the classical regression concept of collinearity.
Since our estimator is a modified regression fit,
it inherits a highly-developed toolbox of 
diagnostics and visualization methods.
Moreover, the proposed method sidesteps
the need for an explicit propensity score model by
instead focusing on the outcome model, which may 
be more attractive in some contexts.
On the other hand, we highlight that the validity
rests on parametric structure, which is not always
appropriate and must be checked carefully. While
IPW-based methods generally also require strong assumptions,
there semiparametric methods that are valid under more general conditions.
In any case, we hope that this work sheds light on 
several interesting and easy-to-compute association 
parameters that
we expect track causal information
more closely than typical regression coefficients.


\section*{Acknowledgements}
We thank Michal Abrahamowicz, Thomas Richardson, Terry Therneau, Lu Tian, and Ang Yu for helpful comments about early versions of this work. R.T.~was supported by the National Institutes of Health (5R01 EB001988-16) and the National Science Foundation (19 DMS1208164). S.B.~was supported by the Foundations of Data Science Institute postdoctoral fellowship.

\bibliographystyle{apalike}
\bibliography{ortho}

\appendix
\newpage

\section{Computation}

We next discuss the computational cost of the nonlinear regression with residuals estimator, and compare it to that of other methods. 
The nonlinear regression with residuals estimator requires $T + 1$ model fits: a model for $X_t$ given the past for $t=1,\dots,T$ and the final regression model with the orthogonalized covariates. Note, however, that if $X_t$ is vector-valued with $d$ coordinates, then fitting the models for $X_t$ given the past may be more involved than a model with a scalar outcome variable. In the extreme case, one could fit each coordinate seperately, which would then amount to $d \cdot T + 1$ total model fits. However, regression models that accommodate a vector-valued responses can be used to keep the total number of model fits at $T + 1$, if necessary. The IPW estimator for the MSM also requires $T+1$ model fits.

We next compare this to the the computation required by the iterative conditional expectations estimator~\citep{hernan2020causal} and the efficient estimator of~\citet{Bang2005}. These two estimators require order $T$ model fits, similar to the nonlinear regression with residuals method. However, these two methods only estimate $\phi(\bar{a})$ for a single pre-specified $\bar{a}$; to estimate $\phi(\bar{a})$ for two different values 
of $\bar{a}$, two different sequences of models must be fit. By contrast, nonlinear regression with residuals and IPW-fitted MSMs give an estimate for $\phi(\bar{a})$ for all $\bar{a}$ in one pass, making them much more efficient when the analyst is interested in understanding the effect of many settings of $\bar{a}$.

\section{Identification conditions}
\label{app:identification}

The causal parameter
$\psi(\overline{a}_t)$
is identified and is given by the $g$-formula
under three conditions:

($\mathghost1$) No interference.
$Y_t(\overline{a}_t) = Y_t$
when $\overline{A}_t = \overline{a}_t$.
This means that someone's outcome depends on their treatment but not
other subjects' treatments.

($\mathghost2$) Positivity.
There is an $\epsilon>0$ such that
$\pi(a_t|h_t) \geq \epsilon$
for all $a_t$ and $h_t$ where
$h_t$ denotes all variables that occur before $A_t$.

($\mathghost3$) No unmeasured confounding.
For every $t$,
$A_t$ is independent of $Y_t(\overline{a}_t)$
given all past variables $H_t$.
This means that the data behave like a randomized experiment
where the randomization probabilities can depend on the past observed variables.

\begin{remark}
Even when the g-formula does not correspond to the causal effect,
it may be a quantity of statistical interest, serving as an adjusted measure of association.
We note that nonlinear regression with residuals is estimating the g-formula even when the three conditions 
above fail to hold.
\end{remark}

\section{Proofs}
\label{app:proofs}

\begin{proof} (Proposition~\ref{prop:ortho_linear_general}) \quad
Let $g_t(\bar{a}_{t-1}, \bar{x}_{t-1})) = \E[X_t \mid \overline{X}_{t-1}, \overline{A}_{t-1}]$.
We start from the g-formula and then manipulate 
the expression as follows:
\begin{align*}
\psi(\overline{a}) &= 
\int\dots\int \left[\alpha + \beta^\top \overline{a}  + \sum_{t=1}^T \lambda_t(x_t - g_t(\bar{a}_{t-1}, \bar{x}_{t-1}))\right] \ 
\prod_t p\left(X_t = x_t | \bar{A}_{t-1} = \bar{a}_{t-1}, \bar{x}_{t-1}\right) dx_1\dots dx_T  \\
&= \alpha + \beta^\top \overline{a}.
\end{align*}
\end{proof}

\begin{proof} (Proposition~\ref{prop:ortho_loglinear}) \quad
We start from the g-formula and then manipulate 
the expression as follows:
\begin{align*}
\psi(\overline{a}) &= 
\int\dots\int \exp\{\alpha + \beta^\top \overline{a}  + \sum_t \lambda_t(x_t - g_t(\bar{a}_{t-1}, \bar{x}_{t-1}))\} \ 
\prod_t p\left(X_t = x_t | \bar{A}_{t-1} = \bar{a}_{t-1}, \bar{x}_{t-1}\right) dx_1\dots dx_T  \\
&= \int\dots\int \exp\{\alpha + \beta^\top \overline{a}  + \sum_t \lambda_t x_t\} \ 
\prod_t p\left(X_t = x_t + g_t(\bar{a}_{t-1}, \bar{x}_{t-1}) | \bar{A}_{t-1} = \bar{a}_{t-1}, \bar{x}_{t-1}\right) dx_1\dots dx_T  \\
&= \int\dots\int \exp\{\alpha + \beta^\top \overline{a}  + \sum_t \lambda_t x_t\} \ 
\prod_t p\left(\epsilon_t = x_t\right) dx_1\dots dx_T  \\
&= \exp\{\alpha + \beta^\top \overline{a}\}\int\dots\int \exp\{\sum_t \lambda_t x_t\} \ 
\prod_t p\left(\epsilon_t = x_t\right) dx_1\dots dx_T  \\
&= \exp\{\alpha' + \beta^\top \overline{a}\}
\end{align*}
for some new intercept $\alpha'$. 
\end{proof}

\begin{proof} (Proposition~\ref{prop:ortho_probit}) \quad
As usual, we will use the g-formula to derive and expression
for $\psi$, and we will relate this to the regression
coefficients from \eqref{eq:probit_assumption}.
\begin{align*}
\psi(\overline{a}) &= 
\int\dots\int \Phi\left(\alpha + \beta^\top \overline{a}  + \sum_t \lambda_t(x_t - g_t(\bar{a}_{t-1}, \bar{x}_{t-1}))\right) \ 
\prod_t p\left(X_t = x_t | \bar{A}_{t-1} = \bar{a}_{t-1}, \bar{x}_{t-1}\right) dx_1\dots dx_T  \\
&= \int\dots\int \Phi\left(\alpha + \beta^\top \overline{a}  + \lambda^\top \overline{x}\right) \ 
\prod_t p\left(X_t = x_t + g_t(\bar{a}_{t-1}, \bar{x}_{t-1}) | \bar{A}_{t-1} = \bar{a}_{t-1}, \bar{x}_{t-1}\right) dx_1\dots dx_T  \\
&= \int\dots\int \Phi\left(\alpha + \beta^\top \overline{a}  + \lambda^\top \overline{x}\right) \ 
\prod_t p\left(\epsilon_t = x_t \right) dx_1\dots dx_T  \\
&= \int\dots\int P(Z \le \alpha + \beta^\top \overline{a}  + \lambda^\top \overline{x})
\prod_t p\left(\epsilon_t = x_t \right) dx_1\dots dx_T  \\
&= P(Z - Z' \le \alpha + \beta^\top \overline{a}) \\
&= P(\sqrt{1 + \sigma^2} Z \le \alpha + \beta^\top \overline{a})
\end{align*}
where $Z \sim \mathcal{N}(0,1)$ and 
$Z' \sim \mathcal{N}(0, \sigma^2)$, independently.
\end{proof}

\begin{proof} (Proposition~\ref{prop:ortho_cox}) \quad
We apply the g-formula to the indicator of the 
event $\{Y \ge t\}$ for each $t$ to compute the 
survival times of the marginal model:
\begin{align*}
\psi(\overline{a}, t) &= 
\int\dots\int \exp\left\{-H_0(t)\exp\{\beta^\top \overline{a}  + \sum_t \lambda_t(x_t - g_t(\bar{a}_{t-1}, \bar{x}_{t-1}))\}\right\} \cdot 
\\ &\qquad\qquad \prod_t p\left(X_t = x_t | \bar{A}_{t-1} = \bar{a}_{t-1}, \bar{x}_{t-1}\right) dx_1\dots dx_T  \\
&= \int\dots\int \exp\left\{-H_0(t)\exp\{\beta^\top \overline{a}  + \sum_t \lambda_t x_t\}\right\} \cdot
\\ &\qquad\qquad \prod_t p\left(X_t = x_t + g_t(\bar{a}_{t-1}, \bar{x}_{t-1}) | \bar{A}_{t-1} = \bar{a}_{t-1}, \bar{x}_{t-1}\right) dx_1\dots dx_T  \\
&= \int\dots\int \exp\left\{-H_0(t)\exp\{\beta^\top \overline{a}  + \sum_t \lambda_t x_t\}\right\} \ 
\prod_t p\left(\epsilon_t = x_t\right) dx_1\dots dx_T  \\
&= \int \exp\left\{-H_0(t)\exp\{\beta^\top \overline{a} + Z\}\right\} \ p(Z = z) dz  
\end{align*}
\end{proof}

\end{document}